\documentclass[preprint]{aastex701}

\newcommand{\CI}{\textit{CI}}

\begin{document}

\small

\title{Intercepting Interstellar Objects\footnote{White paper submitted to the Planetary Science theme of the UK Space Frontiers 2035 exercise.}}

\author[0000-0001-9328-2905]{Colin Snodgrass}
\affiliation{Institute for Astronomy, University of Edinburgh, Royal Observatory, Edinburgh EH9 3HJ, UK}
\email[show]{csn@roe.ac.uk}  

\author[0000-0001-5797-914X]{Marina Galand}
\affiliation{Department of Physics, Imperial College London, London SW7 2AZ, United Kingdom}
\email[]{m.galand@imperial.ac.uk}

\author[0000-0002-5644-2022]{Arnaud Beth}
\affiliation{Department of Physics, Imperial College London, London SW7 2AZ, United Kingdom}
\email[]{a.beth@imperial.ac.uk}

\author[0000-0003-4468-0338]{Charlotte Goetz}
\affiliation{Faculty of Science and Engineering, Northumbria University, Newcastle NE1 8ST, UK}\email[]{charlotte.goetz@northumbria.ac.uk}

\author[0000-0003-4507-9384]{Abbie Donaldson}
\affiliation{Institute for Astronomy, University of Edinburgh, Royal Observatory, Edinburgh EH9 3HJ, UK}
\email[]{a.donaldson@ed.ac.uk}  

\author[0000-0002-9298-7484]{Cyrielle Opitom}
\affiliation{Institute for Astronomy, University of Edinburgh, Royal Observatory, Edinburgh EH9 3HJ, UK}
\email[]{copi@roe.ac.uk}  


\begin{abstract}
\small

We describe how the ESA Comet Interceptor mission, which is due to launch in 2028/29 to a yet-to-be-discovered target, can provide a conceptual basis for a future mission to visit an Interstellar Object. Comet Interceptor will wait in space until a suitable long period comet is discovered, allowing rapid response to perform a fast flyby of an object that will be in the inner Solar System for only a few years; an enhanced version of this concept could realistically provide the first {\it in situ} investigation of a visitor from another star system.
\end{abstract}


\section{Introduction} 


Comets preserve, to a greater or lesser degree, the planetary building blocks of our Solar System; their constituent ices were formed from our Sun's protoplanetary disc. Since then comet nuclei have remained largely inert, stored in cold environments far from the Sun, in the Kuiper Belt or Oort Cloud. The comets we see have only in the last $\le 10^4$ years had their orbits perturbed to bring them into the inner Solar System, where heating by the Sun causes these ices to sublimate, generating the distinctive coma and tails.

In recent decades great progress has been made in understanding these enigmatic objects through remote observation and a series of spacecraft missions (see \citealt{2024come.book..155S} for a review), in particular with the recent ESA mission \textit{Rosetta}. All missions to date have visited short period comets, which have well-defined orbits that allow missions to be rigorously planned far in advance. Short period comets (with orbital periods of ${\sim}6$ years for Jupiter family comets like the \textit{Rosetta} target 67P/Churyumov-Gerasimenko, or even the 76 year period of 1P/Halley) have encountered the Sun many times, resulting in significant (thermal) evolution of the visible surface with each perihelion passage. For example, one of the most striking \textit{Rosetta} results was the discovery that each perihelion passage not only erodes some of the nucleus away as ice sublimates, but that the majority of the dust lifted from the comet by this sublimation actually falls back, rather than being carried away into the tail \citep{Keller2017}. Thick deposits of such fall-back material blanket the northern hemisphere of 67P. In contrast, the southern hemisphere, which sees much stronger summer heating near perihelion, has a very different appearance, gas production,  and composition \citep{2015Sci...347a1044S,2020SSRv..216..102R} -- despite indications that 67P formed from $\sim$ homogeneous building blocks.

Currently, one of the highest priority areas of research in planetary science is to understand  planet formation in the Solar System and to compare it to other exoplanetary systems. To advance in this, in the coming decade, we should focus more efforts in understanding more pristine comets that have not been processed by solar radiation. In particular we need to improve our understanding on how their properties record the processes occurring in the Sun’s protoplanetary disc: What are their compositions and morphologies before they experience significant heating by the Sun? What volatile ices are present in their nuclei when they first approach the Sun? Which of these drive the activity that is now detected at very large distances from the Sun in new comets? To answer these questions we must take full advantage of the new capabilities that will soon be available: more sensitive surveys (e.g., the Vera Rubin Observatory's LSST) that will find new comets at larger  heliocentric distance, powerful new observatories like JWST and ELT to study them, and a new space mission to investigate a pristine comet \textit{in situ},  {\it Comet Interceptor} (\CI).

The next step beyond the detailed investigation of a relatively pristine remnant from our own Solar System's protoplanetary disc will be to compare this with a body that formed elsewhere, to investigate the commonalities and differences between the planet formation process in different places and times in the galaxy. The discovery of the first macroscopic interstellar objects -- comet-like bodies passing through our Solar System that formed around other stars -- has opened a new and exciting window for understanding how planets form \citep{fitzsimmonsreviewcomets3,jewitteseligman}. Each of the three interstellar objects (ISOs) discovered to date (1I/'Oumuamua, 2I/Borisov, 3I/ATLAS) has caused a flurry of activity and excitement in both the scientific community and the general public, and the possibility of investigating one of these objects \textit{in situ} with a spacecraft has been raised each time. While no realistic mission to any of the three ISOs to date has been proposed, the work currently going on to prepare a mission to a new Solar System comet (ESA's \CI)  points the way towards a feasible way to visit an ISO in the coming decades. In this white paper we describe \CI, and how a similar approach could be employed for a future ISO mission.

\section{Comet Interceptor} \label{sec:CI}


\begin{figure}
    \centering
    \includegraphics[width=1\linewidth]{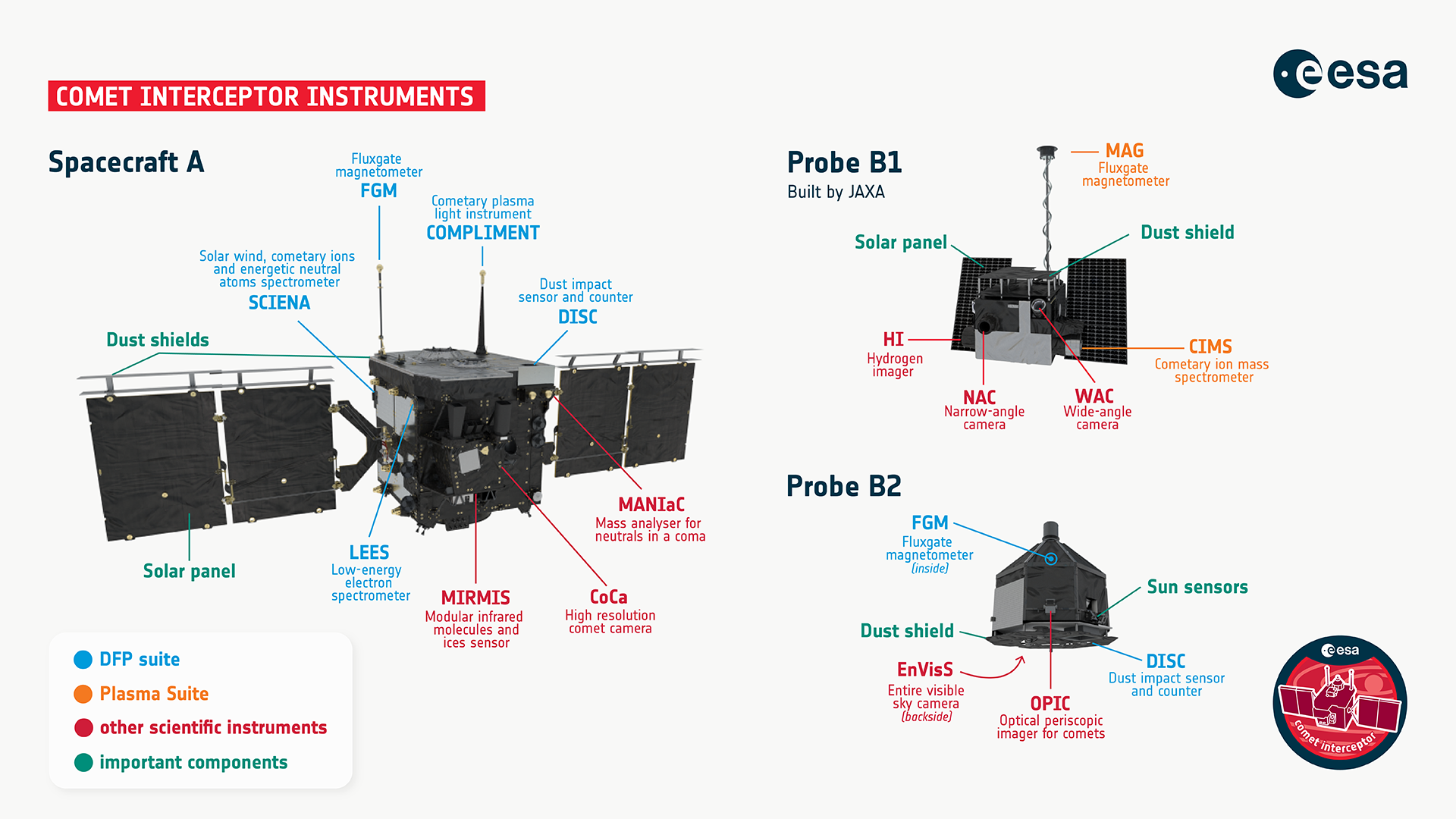}
    \caption{Comet Interceptor spacecraft and instruments (ESA).}
    \label{fig:CIinstruments}
\end{figure}

\subsection{Mission concept}

\CI\ is ESA's first `F-class' mission, a relatively small mission (cost to ESA capped at 150M€) that will launch as a secondary payload. It was originally proposed to launch with the \textit{Ariel} space telescope, but at time of writing alternative launch options are being explored. \CI\ is a multi-spacecraft mission, composed of a primary spacecraft (designated A) and two smaller probes (B1 and B2) that will be released during its comet flyby (Fig. \ref{fig:CIinstruments}). The B1 probe and its instrument payload is provided by JAXA, while spacecraft A and B2 are provided by ESA, with instrument contributions from ESA member states (see section \ref{sec:payload}). 

\begin{figure}
    \centering
    \includegraphics[width=1\linewidth]{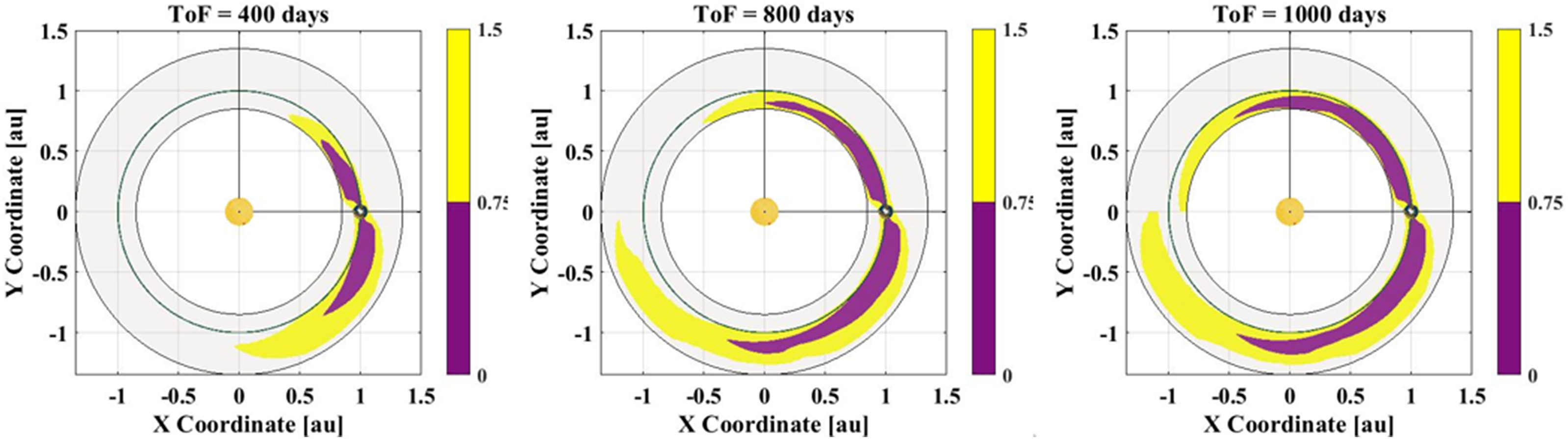}
    \caption{Accessible regions for \CI\ comet encounters, relative to Earth, for different $\Delta v$ (0.75 or 1.5 km/s) and time of flight (ToF) trajectories (from \citealt{2021AcAau.188..265S}).}
    \label{fig:trajectories}
\end{figure}

To get the first detailed look at a primitive comet, \CI\ employs a novel concept to enable it to encounter a long period comet (LPC) approaching the Sun from the Oort cloud, or potentially interstellar space, for the first time: the mission will be designed, built, and possibly launched {\it before} its target is discovered \citep{2019NatCo..10.5418S,2024SSRv..220....9J}. 
\CI\ will launch in 2028/29 and wait in a parking orbit around the Sun-Earth L2 point, where it can station-keep with very little fuel, until a reachable LPC is found. 
The spacecraft will depart L2 to encounter the comet at a distance from the Sun of around 1 au, following a cruise period of up to 3 years. Even with this technique, the target comet must be discovered inbound at a relatively large distance, in order to give sufficient time to characterise its orbit and activity levels, and for the spacecraft to reach the encounter position. Due to spacecraft thermal constraints and the expected amount of available fuel (and therefore $\Delta v$) the encounter must take place at a heliocentric distance $0.9 \le r_h \le 1.2$ au, and close to the ecliptic plane. The volume of space that can be reached within this annulus depends on the available $\Delta v$ and warning time between comet discovery and encounter (Fig.\ref{fig:trajectories}; \citep{2021AcAau.188..265S}).

\subsection{Instrument payload}\label{sec:payload}


\CI\ carries visible-wavelength cameras on all three spacecraft, including narrow and wide angle cameras on probe B1 to study the nucleus and inner coma, respectively, a periscopic imager (OPIC) on B2 that looks forward around the dust shield, and the EnVisS all-sky polarimeter (also on B2) that will study dust reflectance properties. The highest resolution images will come from the CoCa multi-colour camera on spacecraft A, which has telescopic optics based on the CaSSIS instrument on Mars Trace Gas Orbiter \citep{2017SSRv..212.1897T}, to achieve a spatial scale of $\sim$ 10 m/pixel from a nominal minimum flyby distance of 1000 km. The B1 probe also carries a far UV camera that will study the hydrogen coma of the comet (including its D/H ratio) via observation of Lyman-$\alpha$ emission. On spacecraft A direct measurement of molecular and isotopic composition of the gas in the coma will be made by the MANIaC neutral mass spectrometer, which is based on Rosetta/ROSINA heritage \citep{2007SSRv..128..745B}. Further {\it in situ} measurements will be made by the Dust, Field, and Plasma (DFP) package of sensors, which includes magnetometers, dust detectors, particle spectrometers, and electric field and plasma probes dispatched over spacecraft A and B2, and by the Plasma Suite on B1, which is composed of a magnetometer and ion mass spectrometer.

The UK hardware contributions to \CI\ are the FluxGate Magnetometer on probe B2 (FGM-B2) within DFP, and the MIRMIS infrared camera and spectrograph \citep{2020LPI....51.2097B}. The latter is led by the University of Oxford with contributions from VTT Finland and has three channels: near-IR (0.9-1.7 $\mu$m) imaging, mid-IR (2.5-5 $\mu$m) point spectroscopy, and thermal-IR (6-25 $\mu$m) imaging. This will be the first time a thermal-IR instrument is flown to a comet, which will measure temperatures either side of the day/night boundary and provide critical inputs to thermophysical models that explain how cometary activity works. MIRMIS will also measure the composition of both the nucleus surface and the surrounding coma.


FGM-B2 will be measuring the three components of the magnetic field \citep{2023Galand,2024SSRv..220....9J}. It is led by Imperial College London which has a long-standing, well-renowned heritage in space magnetometry, while IWF, Austria, is a key contributor. 
FGM-B2 is composed of two sensors (inboard and outboard) and their electronic boards. 
FGM-B2 aims at characterising the interaction of the supermagnetosonic solar wind with the cometary environment.  Science objectives include the detection of field boundaries (e.g., bow shock, diamagnetic cavity boundary), the identification of their nature, and the assessment of energy and momentum transfer through the cometary environment. For instance, a diamagnetic cavity can build up around the cometary nucleus; the nature of this field-free region is still under debate \citep{2022Goetz}. From low cometary outgassing activity, energy is transferred across the cometary environment through a  variety of waves, to be measured by FGM-B2. Their analysis should shed new light on solar wind-comet interaction. The cometary environment is a rich, natural plasma laboratory that cannot be explored anywhere else within our Solar system. In-situ measurements will provide critical context for remote observations of ions in the coma of comets that cannot be visited.

A key originality of \CI\ is the multi-point capability. The magnetometer is the only {\it in-situ} sensor present on all three spacecraft, while dust densities will be measured at two locations. 
This  three-point measurement capability offers the exciting possibility to characterise  the 3D structure of the magnetic field boundaries for the first time. 


\section{Science requirements for an ISO mission} \label{sec:ISOscience}


ISOs present a unique opportunity to study the building blocks of other planetary systems. It is expected that the planet formation process produces a large number of generally icy small bodies that will be ejected from the system into interstellar space, through gravitational interactions with forming planets. Planet formation models suggest a large fraction of the original bodies in the outer regions of planet-forming discs will be ejected \citep{2020ApJ...904L...4R}, and as a result the galaxy is littered with these objects, which occasionally pass close enough to the Sun and Earth for us to observe. Their properties can give clues as to how planets formed elsewhere in the same way that comets record the process in our own Solar System, and models suggest that the composition of ISOs should vary with source region across the galaxy \citep{2022ApJ...924L...1L,2025ApJ...990L..30H}.

The first ISO observed, 1I/'Oumuamua, was small ($\sim150$m diameter) and surprisingly asteroidal in appearance \citep{2017Natur.552..378M}, despite the expectation that ISOs should be preferentially ejected from the icy outer regions of proto-planetary discs. This meant that it was only discovered as it passed close to Earth, and it was observable for a short period of time ($\sim2$ weeks) before it became too distant and faint to follow. A rapid response from the planetary science community meant that it was as well characterised as possible, with its spectrum \citep[e.g.][]{2018NatAs...2..133F}, lightcurve \citep[e.g.][]{2017ApJ...851L..31K,2017ApJ...850L..36J,2018NatAs...2..383F}, and eventually comet-like non-gravitational acceleration \citep{2018Natur.559..223M} measured. Results are reviewed by the \citet{2019NatAs...3..594O}.

The subsequent ISOs have appeared as comets, which, due to their coma of dust and gas, are much brighter than their inactive nuclei (or asteroids the same size) would be. They were consequently discovered at larger  heliocentric distance and were visible to telescopes for months, despite not coming that close to Earth. Spectroscopic observations revealed bulk coma compositions broadly similar to Solar System comets \citep[e.g.][]{2019A&A...631L...8O,2021A&A...650L..19O,2025MNRAS.544L..31O}, but with some unusual features; 2I/Borisov was CO rich \citep{2020NatAs...4..861C,Bodewits2020}, while 3I/ATLAS (the first ISO to be observable with JWST) has more CO$_2$ \citep{2025ApJ...991L..43C} and an unusual ratio of Fe/Ni \citep{2025arXiv250818382R}.

The next step in ISO exploration is clearly {\it in situ} investigation with a spacecraft -- this is the only way to get resolved images of the nucleus of a comet, and, in the case of active ISO, measurements of detailed gas and plasma composition, and interaction with the space environment. Images and surface composition will reveal clues about planetesimal formation processes around other stars, and the slow evolution of ISOs during their drift through the galaxy. The possibility of direct composition measurements of extra-solar material is very exciting; it can test the universality of ice and mineral content in small bodies, and probe organics, revealing whether or not the chemical building blocks for life are present around other stars.

\section{Mission concepts for an ISO mission} \label{sec:ISOmission}

Papers that describe trajectories for missions to chase after known ISOs \citep[e.g.][]{2022AdSpR..69..402H} serve to demonstrate the impractical nature of such an approach: even if we accept the feasibility of the untested manoeuvres and/or propulsion in these proposals, no consideration is given to how a spacecraft can operate at, communicate from, or even navigate to its faint target, 100 au or more from the Sun. A realistic ISO mission will need to encounter its target during the brief time it is in the inner Solar System, which implies having a spacecraft ready, developed in advance of the discovery of the ISO. One could consider re-tasking an existing spacecraft, although these would not typically have suitable instrumentation for a fast flyby of a comet-like ISO:
for 3I, \citet{2025arXiv250715755Y} found that sending a spacecraft from Mars would require considerably lower $\Delta v$ than a trajectory from the vicinity of Earth, due to the closer approach to that planet.

\citet{2021P&SS..19705137M} studied a mission concept with some similarities to the {\it Deep Impact} comet mission \citep{2005Sci...310..258A}; it would include a flyby spacecraft equipped with a remote sensing payload and a targeted impactor designed to release material from the ISO interior. An interesting feature of this concept was that it was proposed to wait for a target on the ground, with a rapid launch schedule once a suitable target was identified, and a direct launch into an intercept trajectory. In the modern era of frequent commercial launches, such a strategy could be plausible, with a suitable agreement with a launch provider to `skip the queue' when a target is found. \citet{2024P&SS..24105850S} also considered this option, but find that a launch within 30 days of ISO discovery is required to be more effective in reaching ISOs than a \CI-like `wait in space' approach, and reject the ground-storage and rapid launch option as unfeasible for an interplanetary class spacecraft. \citet{abell_2024_y7pqb-fmp74} discuss the various pros and cons of different rapid response strategies in more detail. \citet{2024P&SS..24105850S} propose a single spacecraft with remote sensing instruments based on recent mission heritage, and find that a mission with $\Delta v$ capability of 3 km/s, able to operate out to $r_h$ = 3 au, and perform flybys with relative speed up to 100 km/s, has a reasonable chance of being able to encounter an ISO with a 10 year mission. It can be seen that all of these are scaled up with respect to \CI's capabilities ($\Delta v$ = 0.6 km/s, $r_h <$ 1.2 au, $v <$ 70 km/s, 6 years), but not by unreasonable amounts for a larger class of mission with a dedicated launch.
If \CI\ were already in space 3I would have been beyond its reach, even if it had been discovered much earlier \citep{2025RNAAS...9..207S}, but this ISO was particularly poorly placed for interception, with a perihelion far from Earth: minimum $\Delta v$ trajectories to it also resulted in fast ($\sim 80$ km/s) flybys, while later slower encounters required unfeasible $\Delta v > 70$ km/s \citep{2025arXiv250715755Y}. 
While \CI\ can perform flybys at $10 < v < 70$ km/s, a choice of target that results in as slow an encounter as possible is preferred to maximise the science return \citep{2025arXiv251120521S}.

In terms of payload, owing to the unusual neutral composition that may differ significantly from other Solar System comets, a neutral mass spectrometer is a crucial instrument. Here we differ from \citet{2024P&SS..24105850S}, who consider a mass spectrometer as an optional payload and leave it off of their core payload on cost/complexity grounds: we feel that an ISO mission presents an opportunity to sample extra-solar material that cannot be missed, and that the UK / Europe has the heritage in mass spectrometers \citep[e.g.][]{2007SSRv..128..363W,2007SSRv..128..745B} to achieve this. Some instrument designs may allow to probe ion composition as well but both measurements cannot be performed at the same time by the same device. The most suitable design is a time-of-flight spectrometer to accommodate the short window of observation. 
Performing {\it in situ} plasma observations would be relevant for assessing the solar wind-comet interaction and in support of remote sensing instruments (e.g., UV), for an ISO with outgassing rates not yet encountered in the Solar System ($Q \gtrsim 10^{30}\,\text{s}^{-1}$, affecting the size of the interaction region) and/or very different neutral/ion composition (e.g., mass-loading by heavier ions) and/or different dust level.

A capable remote sensing payload is also essential: resolved imaging will reveal the bulk structure and geomorphology of the nucleus, including evidence for or against craters, layering, and/or substantial surface processing by cometary activity, all of which are critical to understanding the ISO's birth environment around another star. Given the likely high-speed encounter, a long-range telescopic camera will give resolved images over a longer time frame, and can also be useful for navigation. Surface composition can be measured using multi-colour filter imaging or spectroscopy across a range of wavelengths from the UV to the IR; the IR is probably the highest priority for identifying solid-state absorption features in mineralogy and ices. MIRMIS on \CI\ will be the first thermal IR camera on a comet mission, and promises to reveal activity processes -- this would also be valuable for studying an ISO. Assuming a similar spacecraft design to \CI, with dust shields to protect from impacts during a high-speed comet flyby, the critical technology for remote sensing will be precise and rapidly rotating tracking mirrors.


\section{UK priorities for 2025-2035}

The exploration of the small bodies of our Solar System is a high priority area in planetary sciences internationally, and an important one for the UK: in addition to leading the \CI\ proposal, UK-led teams have recently proposed other ESA missions to explore the asteroid belt \citep{2018AdSpR..62.1998B} and the population of comets within it \citep{2018AdSpR..62.1921J,2018AdSpR..62.1947S}, and have played leading roles in developing Near-Earth Object mission concepts for sample return \citep{2009ExA....23..785B} and planetary defence (the Don Quijote Mission -- \citealt{2008LPICo1405.8241W}) that would eventually inspire the current ESA Space Safety programme, and Hera \citep{2025SSRv..221...70M}. \CI\ is an exciting mission concept that will be a highlight of the ESA science programme in the coming decade and is rightly a current UK priority. While \CI\ could potentially visit an ISO if there is a very fortuitous discovery and close approach of an ISO at the right time, this is unlikely. \CI's science goals and mission design are to study a LPC from our own Oort Cloud, which is a high priority area that will advance our understanding of planet formation processes in its own right, but it is notable that the first question asked after any \CI\ talk is inevitably `Could you go to an ISO?'. There is strong scientific and public interest in an ISO mission, and we believe that \CI\ can act as a proof of concept for the `wait in space' approach for such a mission, while achieving its own Solar System exploration goals. 
In the coming decade the LSST (which the UK has also made significant investments in) will likely detect $\sim$ 10 or more ISOs \citep{2022PSJ.....3...71H}, and will place strong constraints on the true population size and therefore likely waiting time to find a reachable one; a necessary step to make a convincing case for a mission.
We have outlined some of the aspects that need to be considered to go from \CI\ to an ISO mission, and believe that further study of these should be a high priority for the international planetary science community in the next decade. The UK space community is well placed to lead such studies, if resourced to do so.

\begin{acknowledgments}
We thank the UKSA and ESA for their support in developing \CI, and the entire \CI\ mission team, past and present for all their contributions.
\end{acknowledgments}

\appendix

List of supporting signatories for this white paper:

\vspace{11pt}
\begin{itemize}
    \item Arnaud Beth, Imperial College London
    \item Beth Biller, University of Edinburgh
    \item Faye Davies, The Open University
    \item Abbie Donaldson, University of Edinburgh
    \item Marina Galand, Imperial College London
    \item Charlotte Goetz, Northumbria University
    \item Jake Hanlon, Mullard Space Science Laboratory, University College London
    \item Pierre Henri, CNRS, Univ Orléans, CNES / Observatoire de la Côte d'Azur
    \item Stephen Lowry, UK Comet Interceptor group
    \item Elena Martellato, INAF - Astronomical Observatory of Padova
    \item Madeleine McLeod, University of Edinburgh
    \item Bob Morris, The Northern Space Consortium 
    \item Brian P. Murphy, University of Edinburgh
    \item Vincent Okoth, University of Edinburgh
    \item Cyrielle Opitom, University of Edinburgh
    \item Maurizio Pajola, INAF-OAPD
    \item Mia Belle Parkinson, University of Edinburgh
    \item Agata Ro{\.z}ek, University of Edinburgh
    \item Martin Rubin, University of Bern
    \item Katherine Shirley, University of Oxford
    \item Colin Snodgrass, University of Edinburgh
    \item Martin D. Suttle, The Open University
\end{itemize}

\bibliography{references}{}
\bibliographystyle{aasjournalv7}

\end{document}